# COUNTERING PRIVACY NIHILISM

*Severin Engelmann\* & Helen Nissenbaum\*\**

Digital Life Initiative, Cornell Tech, New York City

## Abstract

Of growing concern in privacy scholarship is Artificial intelligence (AI), as a powerful producer of *inferences*. Taken to its limits, AI may be presumed capable of inferring "everything from everything," as such, making untenable any normative scheme, including privacy theory and privacy regulation, which rests on protecting privacy based on categories of data—sensitive versus non-sensitive, private versus public. Discarding data categories as a normative anchoring in privacy and data protection as a result of an unconditional acceptance of AI's inferential capacities is what we call *privacy nihilism*. An ethically reasoned response to AI inferences requires a sober consideration of AI capabilities rather than issuing an epistemic carte blanche. We introduce the notion of *conceptual overfitting* to expose how privacy nihilism turns a blind eye toward flawed epistemic practices in AI development. Conceptual overfitting refers to the adoption of norms of convenience that simplify the development of AI models by forcing complex constructs to fit data that are conceptually under-representative or even irrelevant. While conceptual overfitting serves as a helpful device to counter normative suggestions grounded in hyperbolic AI capability claims, AI inferences shake any privacy regulation that hinges protections based on restrictions around data categories. We propose moving away from privacy frameworks that focus solely on data type, neglecting all other factors. Theories like contextual integrity evaluate the normative value of privacy across several parameters, including the type of data, the actors involved in sharing it, and the purposes for which the information is used.

\* Postdoctoral Fellow, Digital Life Initiative at Cornell Tech
\*\* Andrew H. and Ann R. Tisch Professor of Information Science, Cornell Tech at Cornell University.
For their editorial guidance and support, we would like to extend a special thanks to Elisa Orrú and Ralf Poscher at the Max Planck Institute for the Study of Crime, Security, and Law. The authors gratefully acknowledge the valuable feedback on draft versions of this article from participants at the 2024 *Privacy and Public Policy Conference* at Georgetown University, the xD Team at the U.S. Census Bureau, the 2024 *eLaw Conference Law and/versus Technology: Trends for the New Decade* at Leiden University, the 2024 *Conference on Conceptions of Data Protection and Privacy* at the Max Planck Institute for the Study of Crime, Security, and Law, and the *Digital Life Initiative Reading Group* at Cornell Tech. We also extend our sincere gratitude to Daniel Susser, Francien Dechesne, Margot Hanley, Ben Sobel, Kat Geddes, David Widder, and Ben Laufer for their insightful comments on this work. The authors wish to note that co-presence in this volume in no way aligns with other authors, whose views run contrary to fundamental human rights and duty of compassion for all human beings.

## I) INTRODUCTION

Remarkable capacities of artificial intelligence (AI) inferring novel information from information at hand have fueled its growing influence on nearly every aspect of society. The dramatic advancements in inferential analysis based on massive collection, storage, and processing of data have provoked enthusiasm and excitement among AI practitioners and designers.[1] Popular narratives frame AI models to be able to extract nearly any information of interest from nearly any data available.[2] It's not difficult, however, to see how these ascendant inferential powers may evoke its opposite among privacy advocates and regulators. Indeed, regulatory resignation grows in the shadow of AI excitement.[3] Along with AI enthusiasm there is a sentiment of hopelessness in privacy scholarship that AI has become an all-encompassing technology of discovery, so powerful that it may even be capable of inferring "everything from everything" (EfE).[4] Accepting this premise would have profound implications for the role of information in privacy and data protection. In the United States (US), the European Union (EU), Canada, and China, privacy and data protection regulation mandate data processing restrictions based on a classification of information into categories, for example, personal information, financial information, and, in particular, sensitive information.[5] Each data type can be subject to varying levels of protection and legal oversight. However, these bedrock privacy regulations are, *prima facie*, vulnerable to the alleged inferential superpowers of AI, developed to transcend semantic categories and effectively nullify information types as a normative parameter in privacy and data protection law.[6]

---

[1] Chris Anderson, 'The End of Theory: The Data Deluge Makes the Scientific Method Obsolete' (2008) 16(7) Wired Magazine 16-07; Nestor Maslej et al, 'The AI Index 2023 Annual Report' (AI Index Steering Committee, Institute for Human-Centered AI, Stanford University, April 2023) https://aiindex.stanford.edu accessed 30 November 2024.

[2] Viktor Mayer-Schönberger and Kenneth Cukier, *Big Data: A Revolution That Will Transform How We Live, Work, and Think* (Houghton Mifflin Harcourt 2013).

[3] Draper NA, Hoffmann CP, Lutz C, Ranzini G and Turow J, 'Privacy Resignation, Apathy, and Cynicism: Introduction to a Special Theme' (2024) 11(3) *Big Data & Society* 20539517241270663; Draper NA and Turow J, 'The Corporate Cultivation of Digital Resignation' (2019) 21(8) *New Media & Society* 1824-1839; Protecting Privacy Online Begins with Tackling Digital Resignation' (The Conversation, 11 April 2023); https://theconversation.com/protecting-privacy-online-begins-with-tackling-digital-resignation-198979 accessed 30 November 2024.

[4] Daniel J Solove, 'Data Is What Data Does: Regulating Based on Harm and Risk Instead of Sensitive Data' (2023) 118 Northwestern University Law Review 1081; Ohm P and Peppet S, 'What If Everything Reveals Everything?' in Brett Frischmann, Michael Madison, and Katherine Strandburg (eds), *Governing Knowledge Commons* (MIT Press 2016) https://doi.org/10.7551/mitpress/10309.003.0010 accessed 30 November 2024.

[5] See, for example, in the General Data Protection Regulation (GDPR) in Europe: https://commission.europa.eu/law/law-topic/data-protection/reform/rules-business-and-organisations/legal-grounds-processing-data/sensitive-data/ what-personal-data-considered-sensitive_en. Accessed last: May 14, 2024; In the United States' sectoral privacy landscape, for example, the The Health Insurance Portability and Accountability Act (HIPAA) of 1996 specifies 'protected sensitive' data, see: U.S. Department of Health and Human Services, 'Privacy' (HHS.gov) https://www.hhs.gov/hipaa/for-professionals/privacy/index.html accessed 30 November 2024; Canada's Personal Information Protection and Electronic Documents Act (PIPEDA) protects information according to their sensitivity, as well as the amount, distribution, format, and method of storage, see: Office of the Privacy Commissioner of Canada, 'Privacy' (Office of the Privacy Commissioner of Canada) https://www.priv.gc.ca/en/ accessed 30 November 2024.

Different intellectual traditions and disciplines evaluate the legitimacy of inferences through distinct epistemic standards.[7] AI inferences raises fundamental questions of legitimacy: Who determines the standards governing inference, whose interests do these standards reflect, and what justifies their authority? Establishing privacy norms in the context of AI inferences requires careful consideration of the epistemic standards we choose to adopt. Historically, arriving at novel insights or new information through data aggregation, disaggregation, combination, or repurposing—driven by advancements in information technology (e.g., database linking)—could have challenged the normative role of information categories in privacy and data protection. Yet, neither the privacy fears of a "dossier society"[8], voiced as a response to the computerization of records in taxation, health and education in the 1970s and 80s, nor concerns around the construction of comprehensive "mosaics"[9] due to Big Brother-like digital state surveillance in the 2000s resulted in a discontinuation of data categories in data protection law. Unlike these historic vignettes, the current wave of resignatory concerns frames AI as a powerful technology of discovery that moves far beyond a Sherlock Holmesian piecing together of disconnected information to arrive at novel insights. Behind EfE rests a worldview that assumes AI can infer anything of interest about us from anything we reveal, whether intentionally or unintentionally, explicitly or implicitly. *Privacy nihilism*, as we use the term, refers to the abandoning of data categorization in privacy theory and regulation, as a consequence of embracing an EfE position.

---

[6] Solove, *Data is What Data Does* (2023); Paul Quinn and Gianclaudio Malgieri, 'The Difficulty of Defining Sensitive Data – The Concept of Sensitive Data in the EU Data Protection Framework' (2021) 22(8) German Law Journal 1583; Müge Fazlioglu, 'Beyond the Nature of Data: Obstacles to Protecting Sensitive Information in the European Union and the United States' (2019) 46 Fordham Urban Law Journal 271.
[7] Across disciplines, standards for inference are shaped by distinct rules and procedures that govern how conclusions are drawn and validated. A simplified account might suggest that logical inference prioritizes formal validity and consistency; empirical sciences emphasize falsifiability and experimental verification; statistical inference focuses on probabilistic generalization from samples; causal inference relies on counterfactual reasoning to substantiate causal claims; and computational modeling often privileges predictive accuracy as its primary benchmark. In practice, however, each discipline integrates multiple epistemic norms rather than adhering strictly to a single standard. The crucial point is that inference-making, as an epistemic practice—and AI inferences, as a so-called technology of discovery—are fundamentally about interrogating their justification. Whether producing an interpretation, drawing a conclusion, or even generating knowledge, inference demands a valid pathway to warranted belief.
[8] 'Dossier Society', see for example: Kenneth C Laudon, *Dossier Society: Value Choices in the Design of National Information Systems* (Columbia University Press 1986)
[9] 'Mosaic Theory', see for example: David E Pozen, 'The Mosaic Theory, National Security, and the Freedom of Information Act' (2005) 115 *Yale LJ* 628

In this Article we raise objections against privacy nihilism. Across data collection, ground truth manufacturing, and model evaluation, EfE follows a process of epistemic convenience that we anchor on the notion of *conceptual overfitting*. Conceptual overfitting captures the epistemic shortcutting involved in fitting complex constructs of interest to conceptually under-representative or irrelevant data using AI. We question whether this convenient "anything goes" justification for AI inferences warrants abandoning data types as a focal parameter in privacy theory and law. At the same time, we highlight privacy cannot serve its fundamental values to society by sole reliance on the types of data (or information) involved. Theories like contextual integrity assess the normative value of privacy across several parameters, making them better equipped to address the challenges posed by powerful inferential models.

**II) DEFINING PRIVACY NIHILISM**

The rapid proliferation of powerful AI technologies—including predictive and generative models—has intensified challenges in managing their inferential capabilities. Current enthusiasm surrounding AI as an epistemic tool highlights its promise for uncovering hidden insights and complex patterns from vast datasets. Such optimism is widespread among AI practitioners, scientists, and broader societal discourse.

Pausing for a moment to reflect on AI's advancements, and realizing the significant political and economic influence its creators have already gained and are poised to expand further, inspires both amazement and unease. The 2024 Nobel Prize underscored AI's "techno-epistemic momentum".[10] For the first time, the award in Chemistry was given to AI developers of a major tech company. The AI model, called AlphaFold, built on a mature theoretical understanding of protein synthesis[11], is capable to predictively infer the structure of the human proteome.[12] Recently, researchers have employed generative AI to infer models of research participants using data snippets of real participants, suggesting that such methods could replace

---

[10] Hans-Jörg Rheinberger, 'Infra-experimentality: From Traces to Data, from Data to Patterning Facts' (2011) 49 *History of Science* 337; Chemistry Nobel goes to developers of AlphaFold AI that predicts protein structures, October 9, 2024: https://www.nature.com/articles/d41586-024-03214-7
[11] Kathryn Tunyasuvunakool and others, 'Highly Accurate Protein Structure Prediction for the Human Proteome' (2021) 596 *Nature* 590.
[12] Note that this remarkable breakthrough was deeply informed by existing theoretical understanding and empirical knowledge of protein structures developed by scientists over the last century. For commentary, see also: Mel Andrews, 'The Immortal Science of ML: Machine Learning & the Theory-Free Ideal' (2023) preprint https://rgdoi.net/10.13140/RG2.28311.75685 accessed Nov 25, 2024

traditional approaches involving costly human participants.[13] Early research even suggests that large language models could automate the entire scientific process—from generating novel research ideas and writing code, to conducting experiments, visualizing results, drafting complete scientific papers, and simulating peer-review evaluations.[14]

Politically, a powerful tech-centric oligarchy heavily funded the 2024 U.S. election campaign, paving the way for a government demonstrably committed to advancing its techno-libertarian agenda. The fate of society and democracy is increasingly shaped by a techno-utopian ethos—one that sees humanity transacting through unregulated decentralized crypto services, thriving in metaverses, interacting with LLM-driven "agents," and perhaps, one day, even "colonizing" Mars.[15] It is evident that these developments have fostered a pervasive sense of regulatory resignation, as existing frameworks struggle to keep pace with the rapid evolution of sophisticated AI technologies that shape and mediate our participation in social, political, and cultural life.

An illustrative example of the current resignation is the widespread attribution of virtually limitless epistemic power to AI *inferences* across different research disciplines, in political narratives, and popular imagination.[16] AI inferences—the epistemic practice of producing new information from old using AI—play a foundational role in modern digital systems. Inferential models are integral to data mining, data analytics, and data science. A central paradigm in AI mandates datasets to be divided into three parts: a training set *D*, a validation set *V*, and a testing set *T*. An AI model is trained on *D*, its performance assessed on *V*, and its final evaluation conducted on *T*, taken as an unbiased estimate of how well the model's inferences perform in real-world applications. Generally, inductive reasoning governs model development, while a developed model serves as a deductive lens for interpreting novel data.[17] Search engines and social media

---

[13] Park JS, et al., 'Generative Agent Simulations of 1,000 People' (2024) *arXiv preprint* arXiv:2411.10109; Argyle LP, et al., 'Out of One, Many: Using Language Models to Simulate Human Samples' (2023) 31(3) *Political Analysis* 337.
[14] Lu C, et al., 'The AI Scientist: Towards Fully Automated Open-Ended Scientific Discovery' (2024) *arXiv preprint* arXiv:2408.06292.
[15] See, e.g., Marc Andreessen, *The Techno-Optimist Manifesto*, Andreessen Horowitz, Oct. 16, 2023, https://a16z.com/the-techno-optimist-manifesto/; see also Edward Luce, *Beware Elon Musk's Warped Libertarianism*, Financial Times, May 24, 2023, https://www.ft.com/content/7ca8e38d-8ecf-4de2-870d-9bd0a2476147.
[16] Heinrich Peters and Sandra C Matz, 'Large Language Models Can Infer Psychological Dispositions of Social Media Users' (2024) 3(6) PNAS Nexus 231; The Economist, 'What Machines Can Tell from Your Face' (9 September 2017) https://www.economist.com/leaders/2017/09/09/what-machines-can-tell-from-your-face accessed 4 December 2024.
[17] Arvind Narayanan and Sayash Kapoor, AI Snake Oil: What Artificial Intelligence Can Do, What It Can't, and How to Tell the Difference (Princeton University Press 2024); Severin Engelmann and Orestis Papakyriakopoulos, 'Social Media as Classification

platforms use AI to infer thousands of user attributes for profiling—interests, behaviors, relationships, and moral or political views—which they sell to advertisers for targeted marketing.[18] In computer vision, AI models "make sense" of visual data[19] by drawing inferences about them: vision models in robotics[20], self-driving vehicles[21], emotion recognition systems[22], cancer prediction software[23], and beauty apps[24] analyze the semantics of images and videos—billions of which are generated daily across the internet.[25] Every few milliseconds, smart TVs take visual and auditory screenshots to infer viewing habits.[26] On lending platforms AI models infer credit loaners' creditworthiness from both traditional and non-traditional variables.[27] Each year, millions of job applicants undergo video interviews analyzed by AI models inferring employability.[28] Fitness and health trackers gather physiological signals (e.g., heart rate, body temperature) to infer sleep quality and physical activity.[29] Devices such as augmented reality (AR) glasses, virtual reality (VR) headsets, and mixed reality (MR) leverage biometric and neural data to infer their cognitive-emotional content, such as stress and intentions.[30] Large language models (LLMs) are "inference machines," with

---

Systems: Procedural Normative Choices in User Profiling' in Handbook of Critical Studies of Artificial Intelligence (Edward Elgar Publishing 2023) 619–630.
[18] Athanasios Andreou and others, 'Investigating Ad Transparency Mechanisms in Social Media: A Case Study of Facebook's Explanations' in NDSS 2018 - Network and Distributed System Security Symposium (2018); Severin Engelmann et al, 'Social Media Profiling Continues to Partake in the Development of Formalistic Self-Concepts. Social Media Users Think So, Too' (2022) *Proceedings of the AAAI/ACM Conference on AI, Ethics, and Society*.
[19] Dulari Bhatt and others, 'CNN Variants for Computer Vision: History, Architecture, Application, Challenges and Future Scope' (2021) 10 *Electronics* 2470.
[20] Filippo Cavallo and others, 'Emotion Modelling for Social Robotics Applications: A Review' (2018) 15 *Journal of Bionic Engineering* 185.
[21] Joel Janai and others, 'Computer Vision for Autonomous Vehicles: Problems, Datasets and State of the Art' (2020) 12 *Foundations and Trends in Computer Graphics and Vision* 1.
[22] Aya Hassouneh, A M Mutawa and M Murugappan, 'Development of a Real-Time Emotion Recognition System Using Facial Expressions and EEG Based on Machine Learning and Deep Neural Network Methods' (2020) 20 *Informatics in Medicine Unlocked* 100372.
[23] Howard Lee and Yi-Ping Phoebe Chen, 'Image Based Computer Aided Diagnosis System for Cancer Detection' (2015) 42 *Expert Systems with Applications* 5356.
[24] Visage Technologies, 'Visage Technologies' https://visagetechnologies.com/ accessed 30 November 2024
[25] Queensland University of Technology, '3.2 Billion Images and 720,000 Hours of Video Are Shared Online Daily. Can You Sort Real from Fake?' https://www.qut.edu.au/insights/business/3.2-billion-images-and-720000-hours-of-video-are-shared-online-daily.-can-you-sort-real-from-fake#:~:text=QUT%20%2D%203.2%20billion%20images%20and%20720%2C000,daily.%20Can%20you%20sort%20real%20from%20fake accessed 30 November 2024.
[26] Anselmi G, Vekaria Y, D'Souza A, Callejo P, Mandalari AM, and Shafiq Z, 'Watching TV with the Second-Party: A First Look at Automatic Content Recognition Tracking in Smart TVs' in *Proceedings of the 2024 ACM on Internet Measurement Conference* (November 2024) 622-634
[27] For example, Upstart, 'Upstart' https://www.upstart.com accessed 30 November 2024; See also: Mo Chen, Severin Engelmann, and Jens Grossklags, 'Social Credit System and Privacy' in *The Routledge Handbook of Privacy and Social Media* (Taylor & Francis 2023).
[28] Stephen A Woods and others, 'Personnel Selection in the Digital Age: A Review of Validity and Applicant Reactions, and Future Research Challenges' (2020) 29 *European Journal of Work and Organizational Psychology* 64.
[29] Balaji, Ananta Narayanan, and Li-Shiuan Peh, 'AI-On-Skin: Towards Enabling Fast and Scalable On-body AI Inference for Wearable On-Skin Interfaces' (2023) 7(EICS) *Proceedings of the ACM on Human-Computer Interaction* 1-34
[30] Patrick Magee, Marcello Ienca and Nita Farahany, 'Beyond Neural Data: Cognitive Biometrics and Mental Privacy' (2024) 112 *Neuron* 3017.

LLM-based chatbots inferring user preferences, intentions, and goals to act as intimate companions, friends, romantic partners, or mentors.[31]

In a digital economy driven by a data imperative[32], organizations must use AI for inferential techniques to extract value from data through models that infer something new from the data collected. Inferences make data useful, they are the primary value proposition of today's digital economy, more aptly called an "inference economy".[33] One would expect academic scholarship in privacy and data protection to be on high alert, taking a leading role behind the regulatory and policy wheel, steering towards proposals grounded in a careful evaluation of AI inferences.

"**EVERYTHING FROM EVERYTHING**" **(EfE)**

What we observe, however, is a sense of awe tempered by a sense of resignation and hopelessness. In light of grandiose inferential claims published by parts of the AI community, the allure of attributing conclusive epistemic authority to AI inferences is perhaps understandable. Well-known examples include deep learning models that discover depressive markers in the color profile of social media images[34], sexual preferences in people's facial morphology[35], and political orientation in aggregated location data.[36] In these cases, mundane data, easy to amass in today's data economy, are presumed sufficient for AI models to discover sensitive data constructs (e.g., health, sexual and political orientation) from non-sensitive data types. The promise of AI inference-making suggests that nearly *any* data can serve as a potential source of meaning in the process of inferring any construct of interest.[37] This perspective often represents a broader epistemic shift toward seemingly meaningful semantics produced solely based on statistical

---

[31] Iason Gabriel and others, 'The Ethics of Advanced AI Assistants' (2024) arXiv preprint arXiv:2404.16244
[32] Fourcade M and Healy K, 'Seeing Like a Market' (2017) 15 *Socio-economic Review* 9.
[33] Solow-Niederman A, 'Information Privacy and the Inference Economy' (2022) 117 *Northwestern University Law Review* 357.
[34] Andrew G Reece and Christopher M Danforth, 'Instagram Photos Reveal Predictive Markers of Depression' (2017) 6(1) *EPJ Data Science* 15
[35] Yilun Wang and Michal Kosinski, 'Deep Neural Networks Are More Accurate than Humans at Detecting Sexual Orientation from Facial Images' (2018) 114(2) Journal of Personality and Social Psychology 246
[36] Ilaria Liccardi, Alfie Abdul-Rahman, and Min Chen, 'I Know Where You Live: Inferring Details of People's Lives by Visualizing Publicly Shared Location Data' in *Proceedings of the 2016 CHI Conference on Human Factors in Computing Systems* (2016).
[37] Hinds, Joanne, and Adam N Joinson, 'What Demographic Attributes Do Our Digital Footprints Reveal? A Systematic Review' (2018) 13(11) *PLoS One* e0207112; Privacy International, *Data Points Used in Tracking* (Privacy International, 2018) https://privacyinternational.org/sites/default/files/2018-04/data%20points%20used%20in%20tracking_0.pdf accessed 15 December 2024; Kosinski, Michal, David Stillwell, and Thore Graepel, 'Private Traits and Attributes Are Predictable from Digital Records of Human Behavior' (2013) 110(15) *Proceedings of the National Academy of Sciences* 5802-5805

evidence. The presumed revelatory potential of credit in the US offers a compelling case. Insurance companies often predictively correlate a person's financial credit score with their insurance premiums.[38] The underlying rationale is that a person's financial behavior reliably predicts how well they will maintain their home or drive safely. Such inferences are primarily actuarial, where the correlation itself serves as the basis for guiding decision-making procedures. When authority is given to purely statistical evidence, assumptions about the subject matter and a conceptual justification of the relationship between data gathered and constructs inferred typically play a minimal role. Extending this idea further, one might come to believe that ever-increasing datasets and powerful algorithms are able to demonstrate that, after all, "every fact about an individual reveals every other fact about that individual".[39] Data attributes considered sensitive, such as sexual orientation, ethnicity, mental health conditions, or religious beliefs, can be inferred from (nearly) any non-sensitive data proxy such as household electricity patterns, images and videos, mouse clicks, location data, or purchasing habits, privacy scholarship concludes.[40] After all, traces of sensitivity may be embedded in any of these data, ready to be conveniently unearthed by AI's inferential prowess.

If these beliefs were true, traditional regulatory approaches to privacy based on distinctions between sensitive and non-sensitive data would be severely challenged.[41] In the US, EU, Canada, China, and many other countries, information types represent a central normative parameter in privacy and data protection. Regulatory schemes typically specify a set of attributes that fall into the category of sensitive data such as sexual orientation, ethnicity, or religious beliefs. In the United States provisions protecting special information categories include sectoral privacy regulations such as the Health Insurance Portability and Accountability Act (HIPAA)[42] as well as omnibus state laws such as the California Consumer Privacy Act (CCPA)[43] and the Colorado Privacy Act (CPA).[44] In the EU, the General Data Protection Regulation (GDPR) mandates enhanced protections and legal obligations regarding the processing of sensitive data in Article 9.[45] Processing sensitive data requires data processors to conduct a data protection impact assessment (Article 35) and the appointment of a data protection officer (Article 37).[46] Canada's Personal Information Protection and Electronic Documents Act (PIPEDA) employs a tiered approach to data protection, with heightened safeguards in place for sensitive information categories such as health data, financial records, and sexual orientation.[47] China's Personal Information Protection Law (PIPL), implemented in 2021, categorizes certain data types as sensitive, including biometric information, religious beliefs, medical records, location data, and any information pertaining to individuals under the age of 14.[48]

---

[38] Kashmir Hill, 'Carmakers Want to Track Your Driving Habits to Cut Insurance Rates' (11 March 2024) https://www.nytimes.com/2024/03/11/technology/carmakers-driver-tracking-insurance.html accessed 4 December 2024: "*Insurance companies commonly ask for access to a consumer's credit or risk reports, though they are barred from doing so in California, Massachusetts, Michigan and Hawaii.*"
[39] Ohm and Peppet, *What If Everything Reveals Everything?* (2016)
[40] Solove, *Data is What Data Does* (2023)
[41] Fazlioglu, *Beyond the Nature of Data* (2019); Quinn and Malgieri, *The Difficulty of Defining Sensitive Data* (2021)
[42] Health Insurance Portability and Accountability Act of 1996, Pub L No 104-191, 110 Stat 1936 https://www.govinfo.gov/content/pkg/PLAW-104publ191/pdf/PLAW-104publ191.pdf accessed 2 January 2025.

These regulatory approaches construct a categorical framework in which certain types of information, classified as sensitive, for example, warrant special and restrictive treatment.[49] However, the acceptance of the EfE thesis—the belief that "everything can be inferred from everything"—fundamentally challenges this approach. In summary, there is a prevailing sense of resignation in privacy and data protection, epitomized by EfE, that AI's seemingly boundless inferential capabilities make it impossible to maintain robust privacy norms in theory and regulation.

**III) CONCEPTUAL OVERFITTING**

We contend that privacy nihilism is untenable, as its premise, here defined as EfE, lacks sufficient justification to abandon data types (sensitive, financial etc.) as a normative parameter in privacy and data protection. Nonetheless, we are not oblivious to the many eye-catching illustrations that purportedly claim

---

[43] California Office of the Attorney General, 'California Consumer Privacy Act (CCPA)' (OAG California) https://oag.ca.gov/privacy/ccpa accessed 2 January 2025.
[44] Colorado Attorney General, 'Colorado Privacy Act' (Colorado Attorney General) https://coag.gov/resources/colorado-privacy-act/ accessed 2 January 2025.
[45] European Commission, 'What Personal Data is Considered Sensitive?' (European Commission) https://commission.europa.eu/law/law-topic/data-protection/reform/rules-business-and-organisations/legal-grounds-processing-data/sensitive-data/what-personal-data-considered-sensitive_en accessed 2 January 2025. Accessed last: May 14, 2024.
[46] Regulation (EU) 2016/679 of the European Parliament and of the Council of 27 April 2016 on the Protection of Natural Persons with Regard to the Processing of Personal Data and on the Free Movement of Such Data (General Data Protection Regulation) [2016] OJ L119/1 https://eur-lex.europa.eu/legal-content/EN/TXT/PDF/?uri=CELEX:32016R0679 accessed 2 January 2025.
[47] https://www.priv.gc.ca/en/
[48] Qian Li, Tao Jiang, and Xijian Fan, 'Examining Sensitive Personal Information Protection in China: Framework, Obstacles, and Solutions' (2023) 58(3) *Information & Culture* 247–273; China Briefing, 'The PRC Personal Information Protection Law: Final A Full Translation' (China Briefing) https://www.china-briefing.com/news/the-prc-personal-information-protection-law-final-a-full-translation/ accessed 2 January 2025.
[49] The GDPR justifies sensitive data types also as an "end in itself" in the recognition of their protection as fundamental rights (Quinn & Malgeri, 2021).

AI can infer sensitive from non-sensitive data types. The underlying logic of privacy nihilism, as we have defined it, starts with the premise that AI has virtually limitless capacities to derive new information from information at hand. This position risks being more of a caricature than a defensible stance. To argue this point, we develop the notion of *conceptual overfitting*. Conceptual overfitting documents norms of convenience in producing inferences that seemingly allow fitting complex constructs to conceptually under-representative or irrelevant data. Conceptual overfitting exposes forms of epistemic shortcutting whereby EfE models offer seemingly conclusive statistical evidence but sidestep any meaningful commitment to meet conceptual expectations of contested inference constructs—including political affiliation, sexual orientation, or mental health conditions. Specifying *conceptual* commitments prior to model development would disrupt the frictionless design of models that claim to infer everything from everything. This tendency toward inferential model design reflects broader norms of epistemic convenience, which we examine across three key stages of AI model development: data collection, ground truth manufacturing, and model evaluation.

**DATA COLLECTION**

Contemporary organizations are situated within a data economy that prioritizes data-centric business models and a technological infrastructure engineered to minimize resistance in data collection. Data is abundant, on tap, lying around on easily accessible large public datasets such as ImageNet, LAION-5B, LibriSpeech or CommonCrawl, a publicly available crawl of the web.[50] Even when organizations do not know exactly for what purposes they should collect data, they believe they will lose their place in the data economy if they stop gathering data or delete their data repositories.[51] Powerful and easily accessible, open source data collection tools such as scraping and crawling software require little technical competence to extract troves of data from public and semi-public websites. Organizations that need to fill up their own data repositories pay for data at specialized data gathering companies, data brokers, such as Acxiom or LexisNexis that pull together data from different channels and sources including digital

---

[50] OpenAI's large language models are trained on diverse datasets, including publicly available resources such as The Common Crawl, a comprehensive web crawl dataset, see: Will Knight, 'The Fight Against AI Comes to a Foundational Data Set' *Wired* (13 November 2023) https://www.wired.com/story/the-fight-against-ai-comes-to-a-foundational-data-set/ accessed 8 December 2024.
[51] Sadowski, Jathan. "When data is capital: Datafication, accumulation, and extraction." *Big data & society* 6.1 (2019): 2053951718820549.

traces, exhaust or data dumps. In the prevailing culture of datafication and inference-making, AI-producing organizations are driven by incentives to exploit readily available datasets, extracting value by developing models that generate inferences of interest. This focus on value extraction, underpinned by a "more data is always better"[52] directive, has fostered norms of indiscriminate data amassing to find valuable insights in "data kitchen sinks." A tale known as the Drunkard's Search might serve as an illustration of this phenomenon:

*While on duty, a police officer encounters a drunk individual rummaging around under a streetlight. The officer asks what the person is searching for, and the individual mentions that they have lost their keys. Together, they search beneath the light for a few minutes. Eventually, the officer questions the person's certainty about losing the keys there, prompting the individual to admit they were actually lost in the park. Puzzled, the officer inquires why they are searching in that spot, to which the individual responds: "Because it's illuminated here."*[53]

Organizations are motivated to look for value in the data that they happen to have at their disposal or that they can access with ease. One such data comfort zone has been social media data. Offering semi-public access to the profiles, behaviors, and interactions of billions of individuals globally, social media platforms run on the promise to provide actionable insights on specific, granular audiences to marketers. Data on social media have been a playground for developing and testing inferential AI models. Take for instance the numerous example cases that purportedly infer contested, dynamic constructs such as *political orientation* from social media kitchen sink data.[54] The typical approach involves collecting all accessible profile data—such as likes, number of friends, and other attributes—from users who explicitly disclose their political affiliation. During training, an AI model learns to identify patterns in this kitchen sink dataset and associates them with political affiliations (e.g., "Democrat" or "Republican"). In testing and validation, the model predicts political orientation based solely on the profile data. When deployed, the AI model infers political orientation of users who have not explicitly disclosed it, relying only on their kitchen sink profile data. Eventually, a model appears to discover a dynamic, fluid, contested, and sensitive data type—political orientation—in mundane, conceptually unrelated pieces of information[55] based on the correlation of similarity to users that did reveal this construct (perhaps by accident). AI models are powerful at picking up regularities and patterns in data. But patterns should not be mistaken for semantics when they cannot be evaluated against any conceptual claims about the phenomenon of interest (e.g., political orientation) prior to data collection. The ease of amassing data motivates organizations to bypass data collection norms grounded in explicit assumptions about a construct—and the accountability that comes with committing to them. Tempted by the convenience of data amassing, inferential models fit constructs onto conceptually under-representative and irrelevant data. By sifting through vast quantities of data, it is presumed that contested constructs, such as political or sexual orientation, simply emerge from the data and are ultimately "discovered" by AI models: "Data speak for themselves," so the adage, they can be harvested as intrinsic beares of semantic meaning. Data are, in fact, not factual, fixed representations, they are not gateways to an objective reality. Even if datasets expand to gigantic Internet-sized repositories, their representational scope does not necessarily broaden to the point where AI models can simply pick up nearly any construct of interest, as suggested by privacy nihilism. Conceptual assumptions about the construct of interest guide the search for construct-relevant data in inferential AI models. The notion "accurate data" is meaningless if devoid of a conceptual reference. Against what else should we evaluate the accuracy of our data, if not conceptual claims and assumptions about the constructs we aim to infer? A commitment to accurate data necessitates conceptual commitments prior to data collection. This, in turn, requires discipline-specific expertise and with it a willingness to step outside of data comfort zones—challenging the convenience of data collection driven by a Drunkard Search principle.

---

[52] Such and similar convictions seem widespread in the field of AI. A recent analysis of introductory AI courses highlighted a concerning trend: most courses overemphasized the role of large datasets while neglecting the critical importance of data quality, guided by conceptual assumptions, see: Severin Engelmann, Mihir Z Choksi, Audrey Wang, and Casey Fiesler, 'Visions of a Discipline: Analyzing Introductory AI Courses on YouTube' in *The 2024 ACM Conference on Fairness, Accountability, and Transparency* (ACM, June 2024) 2400–2420.
[53] Right Attitudes, 'Drunkard Search and the Streetlight Effect' (26 February 2016) https://www.rightattitudes.com/2016/02/26/drunkard-search-streetlight-effect/ accessed 30 November 2024.
[54] Boutet A, Kim H and Yoneki E, 'What's in Twitter, I Know What Parties Are Popular and Who You Are Supporting Now!' (2013) 3 *Social Network Analysis and Mining* 1379;  Kosinski M, Stillwell D and Graepel T, 'Private Traits and Attributes Are Predictable from Digital Records of Human Behavior' (2013) 110(15) *Proceedings of the National Academy of Sciences* 5802; Wang Y, Weber I and Mitra P, 'Quantified Self Meets Social Media: Sharing of Weight Updates on Twitter' in *Proceedings of the 6th International Conference on Digital Health Conference* (2016).

Besides, fitting sensitive constructs to data without accountability to any conceptual relevance while hoping all noise will come out in the wash of ML algorithms increases the likelihood of measurement errors and well-documented downstream harms, such as discrimination and disparate treatment—issues that a significant part of the data science community has identified and warned against.[56] It would be puzzling if privacy and data protection norms should promote epistemic standards that reinforce such established risks.

---

[55] Helen Nissenbaum, 'Contextual Integrity Up and Down the Data Food Chain' (2019) 20(1) Theoretical Inquiries in Law 221-256.

**GROUND TRUTH MANUFACTURING**

All supervised AI models require a ground truth—an epistemic reference from which models interpret new data. In the previous example on self-revealed political orientation on social media, the ground truth was conveniently embedded in the collected data as an incidental byproduct. However, in other cases, developers first need to add their construct of interest to the data, so that a model can learn the association between data and inference construct. In other words, developers need to *manufacture* a ground truth. Manufacturing a conceptually justified ground truth can be the foundation for the development of breakthrough inferential AI models such as AlphaFold, the 2024 Nobel Prize-winning AI model for proteome prediction. AlphaFold was trained on high-quality, labeled protein structures data[57] donated to the open source Protein Data Bank (PDB)[58] by over 50,000 structural biologists worldwide starting in 1971.[59] Clearly, such a highly conceptualized, structured, and systematized dataset, built over decades, fundamentally contradicts the notion of convenient inferential model development promoted by EfE.

*Data labeling & annotation*

The PBD demonstrates that the assumption that AI models infer constructs by simply "letting the data speak" collapses since human annotators—whether volunteers, crowdsourced workers, or domain experts—carry out the interpretative task of assigning a target construct, the intended inference concept,

---

[56] Sambasivan N and others, '"Everyone Wants to Do the Model Work, Not the Data Work": Data Cascades in High-Stakes AI' in *Proceedings of the 2021 CHI Conference on Human Factors in Computing Systems* (2021); Wang A and others, 'Measuring Representational Harms in Image Captioning' in *Proceedings of the 2022 ACM Conference on Fairness, Accountability, and Transparency* (2022); Jacobs AZ and Wallach H, 'Measurement and Fairness' in *Proceedings of the 2021 ACM Conference on Fairness, Accountability, and Transparency* (2021); Ullstein C, Engelmann S, Papakyriakopoulos O, Ikkatai Y, Arnez-Jordan NP, Caleno R, ... & Grossklags J, 'Attitudes Toward Facial Analysis AI: A Cross-National Study Comparing Argentina, Kenya, Japan, and the USA' (2024) The 2024 ACM Conference on Fairness, Accountability, and Transparency 2273-2301.
[57] Tunyasuvunakool et al., *Highly Accurate Protein Structure Prediction for the Human Proteome* (2021). We note that AlphaFold not only trained on labelled data in the PBD: *"The AlphaFold architecture is able to train to high accuracy using only supervised learning on PDB data, but we are able to enhance accuracy (Fig. 4a) using an approach similar to noisy student self-distillation."*
[58] RCSB Protein Data Bank, https://www.rcsb.org/ accessed 17 December 2024
[59] Noé F, 'How AlphaFold Works' (Dale on AI, 2024) https://daleonai.com/how-alphafold-works accessed 17 December 2024

to individual data records for model training. Model training crucially relies on the inherently commercial, social, and collaborative process of assigning semantics to data. Organizations typically provide their data either to in-house labelers or to a data processing company to facilitate dataset labeling.[60] Data processing companies supply the necessary infrastructure, including project supervisors, human labelers, and quality control systems, making the process convenient for the requesting organization. Human labelers, however, must navigate the inherent semantic ambiguity, fluidity, procedurality, and contextuality that mediate the relationship between data and inference. Labeling and annotation practices prioritize the pragmatic interests of requester organizations, smoothing over epistemic complexities for convenience.[61] Take, for example, cases where data requesters ask human labelers to assign race and ethnicity categories to people depicted in visual images. Annotation companies provide human labelers with a *predefined* set of race categories such as—"White," "Hispanic," "African American," "Asian," "Indian," or "Ambiguous"—and instruct them to map these constructs onto clients' image data.[62] Race and ethnicity are complex constructs not receptive to self-evident, intuitive identification, unlike distinctions between, say, a tennis ball and a basketball. With no conceptual interpretation in hand, annotators' cultural background and subjective experiences with people of different races and ethnicities influences how they perceive and classify each image into predefined race categories. As a result, disagreement among labelers is a common occurrence. To identify this disagreement, data processing companies invest in 'consistency monitoring,' reviewing labeling patterns and re-training labelers for specific tasks.[63] The most prevalent method for resolving annotator disagreements is majority voting, which creates a ground truth that implies a single correct interpretation for every data record.[64] Each semantic unit of a data record (e.g., a specific part of an image) is labelled by several annotators but can only receive a single semantic labeling. Labeling does not allow for a plurality of semantic annotation. Not only does such annotation bias generate misclassifications that harm data subjects for whom the model incorrectly assigns the majority

---

[60] Muller and others, 'Designing Ground Truth and the Social Life of Labels' in Proceedings of the 2021 CHI Conference on Human Factors in Computing Systems (2021) 1–16.
[61] Severin Engelmann and others, 'What People Think AI Should Infer from Faces' in Proceedings of the 2022 ACM Conference on Fairness, Accountability, and Transparency (2022); Milagros Miceli, Julian Posada and Tianling Yang, 'Studying Up Machine Learning Data: Why Talk About Bias When We Mean Power?' (2022) 6 Proceedings of the ACM on Human-Computer Interaction GROUP 1.
[62] Miceli et al., *Studying Up Machine Learning Data* (2022)
[63] Melchior D, Mistry K and Proskudin K, *Accuracy Upfront: Methods and Process for Measuring and Improving Accuracy in AI through Human Supervision* (CloudFactory Whitepaper, 2023).
[64] Eve Fleisig et al, 'When the Majority is Wrong: Modeling Annotator Disagreement for Subjective Tasks' (17 March 2024) arXiv:2305.06626v5.

interpretation, it risks oversimplifying nuanced and contested constructs including race and ethnicity.[65]

When we hear AI slogans of inferential mastery we should remind ourselves that such epistemic performativity may rest on the meaning that human labelers ascribe to data during socio-commercial negotiations.[66] Privacy nihilism ignores that AI models infer complex constructs such as racial or ethnic categories, emotional profiles, or gender types only because, in a largely invisible infrastructure, a human labeler fits a computationally digestible and readable representation of an infinitely more complex construct to each training data record.

*Survey questionnaires*

Conceptual overfitting occurs when models correlate survey questionnaire results with kitchen sink datasets to infer complex constructs such as medical conditions. A prominent example is AI inferences about mental health, where diagnoses depend on clinicians' interpretation of survey responses in conjunction with other sources of diagnostic evidence.[67] Medical conditions like depression, anxiety, schizophrenia, or bipolar disorder, which each have a unique symptom profile, mental health survey questionnaires play a *supportive* role in screening, diagnosis, or evaluating symptom intensity.[68] Numerous clinically approved mental health questionnaires are available, many of which can be downloaded for free. These questionnaires often include a scoring system that seemingly facilitates straightforward interpretation of results into definite diagnostic categories. Given their availability and apparent ease of interpretation, mental health questionnaires are common tools of ground truth manufacturing for AI models that purportedly infer mental health conditions by correlating questionnaire results with easily available, but conceptually unrelated data.[69] A typical machine learning project might recruit online crowdworkers and request them to complete a clinically-certified survey questionnaire focused on a specific health condition. After completing the task, crowdworkers consent to share unrelated data, such as their social media information. This social media data is then correlated with participants' questionnaire responses to train a model that allegedly infers medical conditions from social media activity.

---

[65] Muldoon, James, Mark Graham, and Callum Cant. "Feeding the Machine The Hidden Human Labour Powering AI." (2024).
[66] Muller M, et al, 'Designing Ground Truth and the Social Life of Labels' in *Proceedings of the 2021 CHI Conference on Human Factors in Computing Systems* (2021); Similar to human labeling in these supervised learning procedures, note that in reinforcement learning from human feedback (RLHF), collecting human preferences over pairs of outputs has become the standard approach for aligning large language models (LLMs).
[67] Shatte ABR, Hutchinson DM, and Teague SJ, 'Machine Learning in Mental Health: A Scoping Review of Methods and Applications' (2019) 49 *Psychological Medicine* 1426.
[68] Shatte et al., *Machine Learning in Mental Health* (2019)
[69] Chancellor S and De Choudhury M, 'Methods in Predictive Techniques for Mental Health Status on Social Media: A Critical Review' (2020) 3(1) *NPJ Digital Medicine* 43; IBID Andrew G Reece and Christopher M Danforth, 'Instagram Photos Reveal Predictive Markers of Depression' (2017) 6(1) *EPJ Data Science* 15; De Choudhury M and others, 'Predicting Depression via Social Media' in *Proceedings of the International AAAI Conference on Web and Social Media* (2013) 7(1).

To give one example, a well-cited study paid online crowdworkers to complete the clinically validated Center for Epidemiologic Studies Depression Scale.[70] Based on their questionnaire responses, researchers classified crowdworkers as either depressed or healthy. Using an AI model, they then analyzed how these classifications correlated with the color hues of crowdworkers' social media images[71]. The authors claim that their random forest classifier identified markers of depression by detecting variations in color hues within social media images of individuals classified as either "depressed" or "healthy." The authors concluded that "markers of depression" could be predictively inferred from social media images. When diagnosing depression, clinicians combine multiple diagnostic methods, including various physical examinations, medical history, and patient interviews to explore symptoms and potential causes (e.g., family history), *as well as* administer depression scales[72] for diagnostic evaluation.[73] Given the complexity of mental health conditions—recognized as sensitive data under most data protection laws worldwide—doctors rely on evidence from a combination of these practices to arrive at a diagnosis. Are we to accept that an AI model inferring depression markers from image color hues captures the constitutive dimensions of depression as comprehensively as trained clinicians who conduct and interpret a range of diagnostic assessments?

When AI projects claim to achieve the same diagnostic value as trained clinicians while relying on conveniently produced ground truths, they engage in conceptual overfitting—reducing a multifaceted construct to conceptually unrelated, surface-level correlations, such as color hues in images, while disregarding its nuanced psychological, social, and biological dimensions. By prioritizing proxies over underlying causes, these AI models risk producing contextually fragile predictions that lack the

---

[70] Reece and Danforth, *Instagram Photos Reveal Predictive Markers of Depression* (2017)
[71] Crowdworkers had consented to sharing their social media images as part of the study.
[72] Examples include the Hamilton Depression Rating Scale, the Montgomery-Åsberg Depression Rating Scale, or the Center for Epidemiologic Studies Depression Scale.
[73] The process of diagnosing depression at a hospital is much more complex than outlined here. For the purpose of the argument, I have drawn information from: https://nyulangone.org/conditions/depression/diagnosis

robustness[74] required for reliable diagnostic applications.[75]

*Proxy Hopping*

In some cases, the inferential leap from data to inferred construct is so large, conceptually, that the construct of interest is first inferred from another construct, itself an inference. For example, AI models take as input signal streams from data primitives such as accelerometer or gyroscope sensors in smart devices and output constructs such as emotions[76], health conditions[77], stress[78], or gender[79] as inferences. We call this *proxy hopping*, as it requires "hopping" from data proxy to data proxy to arrive, finally, at the construct of interest. This recursive cycle of leaping between data and inference necessitates producing a corresponding ground truth for each data-inference pair. In Olsen and Torresen (2016)[80], for example, accelerometer[81] data streams serve as the first proxy $P_1$ to infer a specific type of movement (e.g., walking) $I_1$. $I_1$ is then considered a proxy ($P_2$) for the final inference, emotion ($I_{final}$). To manufacture the ground truth for $P_1 \rightarrow I_1$ participants install a smartphone application that collects accelerometer data while they perform the desired movement (walking, running, jumping etc.). This results in a ground truth of accelerometer data that signifies a particular movement profile. The same participants then complete a survey questionnaire to capture their emotional state[82] in relation to the type of movement ($P_2 \rightarrow I_{final}$). Hence, an AI model makes inferential leaps from sensor streams to emotional states. Since each inference can be mapped onto conceptually irrelevant data, proxy hopping is particularly susceptible to conceptual overfitting. In turn, models that engage in proxy hopping are especially prone to generating downstream harm through misclassification, as measurement errors may compound across inferential leaps, with each error undermining the validity of all subsequent inferences.

---

[74] For example, in Sendhil Mullainathan and Ziad Obermeyer, 'Does Machine Learning Automate Moral Hazard and Error?' (2017) 107(5) American Economic Review 476: "*Machine learning algorithms excel at predicting outcomes (y) based on inputs (x). In automation tasks, measuring y—such as the majority opinion of ophthalmologists—is straightforward. However, in health policy applications, y and x are often derived from electronic health records or claims data, where the very construction of these datasets introduces large and systematic mismeasurement. These inaccuracies, in turn, can bias algorithmic predictions and, in some cases, automate policies that amplify existing clinical errors and moral hazard.*"
[75] For example in Reece and Danforth, *Instagram Photos Reveal Predictive Markers of Depression* (2017), the authors state: "*These results suggest new avenues for early screening and detection of mental illness.*"
[76] Andreas Fsrøvig Olsen and Jim Torresen, 'Smartphone Accelerometer Data Used for Detecting Human Emotions' (2016) 3rd International Conference on Systems and Informatics (ICSAI) (IEEE 2016).
[77] Ahmad Jalal et al, 'A Study of Accelerometer and Gyroscope Measurements in Physical Life-Log Activities Detection Systems' (2020) 20 Sensors 6670.
[78] Enrique Garcia-Ceja, Venet Osmani and Oscar Mayora, 'Automatic Stress Detection in Working Environments from Smartphones' Accelerometer Data: A First Step' (2015) 20 IEEE Journal of Biomedical and Health Informatics 1053.
[79] Ankita Jain and Vivek Kanhangad, 'Investigating Gender Recognition in Smartphones Using Accelerometer and Gyroscope Sensor Readings' (2016) International Conference on Computational Techniques in Information and Communication Technologies (ICCTICT) (IEEE 2016).
[80] Olsen AF and Torresen J, 'Smartphone Accelerometer Data Used for Detecting Human Emotions' (2016) 3rd International Conference on Systems and Informatics (ICSAI), IEEE
[81] A smartphone's accelerometer is a sensor that detects movement and orientation by measuring acceleration along three axes. This allows the device to interpret motion, adjust screen orientation, and enable various interactive features
[82] Olsen and Torresen, *Smartphone Accelerometer Data Used for Detecting Human Emotions* (2016). The emotion questionnaire administered in this study was based on the Circumplex Model of Affect.

Following the easy collection of data, model development often relies on ground truth manufacturing practices that operationalize conceptual overfitting. This occurs when data are assigned the meaning of complex constructs through convenient labeling processes or are merely correlated with questionnaire results or a combination thereof, as is the case with proxy hopping. In fact, human discretionary work infuses model development even before ground truth production: data cannot be labelled, annotated or otherwise humanly interpreted without prior data preparation efforts that require an organizational infrastructure to enable human decision-making on data processing and cleaning.[83] Before a single data record can be semantized by human interpretation for ground truth manufacturing, data must first be made intelligible for interpretative work through filtering (excluding outliers, e.g.), sorting, ordering and a plethora of other human-led practices crucial for ground truth manufacturing. AI models do not uncover constructs of interest from the intrinsic structures and patterns of data. Instead, their representational value is assigned through human modularization, collaboration, and interpretation.[84] When this assignment of meaning prioritizes pragmatic convenience over conceptual rigor, models can conceptually overfit a complex construct onto conceptually under-related or irrelevant data.

**MODEL EVALUATION**

AI models and the inferences they produce are conceptualized, developed, and validated through the dominant train-test paradigm in supervised machine learning. This paradigm mandates that accuracy metrics serve as standard evaluation and validation devices of AI model performance. Privacy nihilism assumes that high accuracy scores are sufficient to declare a model is able to infer one thing from another. However, accuracy rates are empty validators when they are disconnected from any meaningful interpretation about the relationship between the inferred construct and the statistical results. This is particularly true for models that conceptually overfit, where data is collected without a commitment to conceptual assumptions, enabling the manufacturing of ground truth based on oversimplified constructs. Now, at the model evaluation stage, all we have are accuracy metrics that appear convenient to interpret: the higher the better.[85] But how high is high enough? 55%? 78%? 95 or 99%? The reigning heuristic remains: the higher the accuracy score, the better the performance. But no one really seems to be able to tell (including us) when accuracy scores are high enough. Should we deem an AI model to work when it predicts a person's political orientation (Democrat or Republican) with 72% accuracy on a validation set, based solely on facial morphology?[86] Probability scores or percentages lend model evaluation an appearance of scientific objectivity.[87] In the end, however, human designers and practitioners choose an accuracy cut-off threshold that turns a probabilistic expression into a deterministic declaration of classification. For example, if a model predictively infers an 80% probability that an individual should be assigned an inference construct, a cut-off value of 79% would indicate that a relationship between the individual and the inference construct exists. In contrast, a cut-off value of 81% would denote no existing relationship. Accuracy cut-off choices are essentially arbitrary in models that conceptually overfit. Accuracy rates may be persuasive devices of performativity and indicate whether inferential models are "hitting or missing" target constructs.[88] When models conceptually overfit, they cannot be interpreted in relation to any meaningful theoretical claims or assumptions about the target construct. As a result, there is no conceptual accountability for the accuracy they produce.

---

[83] For an in-depth account on the epistemological significance of data cleaning in contemporary data science, see: Marcel Boumans and Sabina Leonelli, *From Dirty Data to Tidy Facts: Clustering Practices in Plant Phenomics and Business Cycle Analysis* (Springer International Publishing 2020).
[84] Florian Jaton, 'Assessing Biases, Relaxing Morality: On Ground-Truthing Practices in Machine Learning Design and Application' (2021) 8 *Big Data & Society* 20539517211013569.

Moreover, since accuracy metrics are often produced by the same actors who stand to benefit from their positive presentation, privacy scholarship should exercise caution in relying on them to inform normative claims for governance proposals. For example, Birhane et al. (2022) show that tests of generalizability usually occur on datasets and tasks that researchers select and have access to.[89] In their analysis of 100 highly cited machine learning papers, they underline the convenience that performance metrics rarely extend beyond curated datasets. Reporting AI accuracy rates under conditions comparable to a laboratory setting is akin to evaluating performance in an overly controlled environment. This practice risks creating an illusion of performance, obscuring the documented challenges and limitations of deploying AI in real-world contexts.[90] Indeed, we observe a growing "how did we get here?" sense of self-reflection among data scientists, which goes hand in hand with the realization that the persistent clinging to accuracy measures might lead the field into a dead end.[91] Current evaluation practices center on accuracy optimization competitions separating winners from losers on leaderboards. This sole focus on accuracy metrics may, ironically, devalue such metrics as a reliable measure of AI performance. Rachel Thomas, an AI scholar, and David Uminsky, an applied mathematician, describe the current evaluation culture in AI—the chase for state-of-the-art (SOTA) accuracy metrics in competitions—by reference to Goodhart's law: "When a measure becomes a target, it ceases to be a good measure".[92] Accuracy metrics serve as sophisticated organizational tools in AI research and practice, exemplified by the prevailing culture of SOTA evaluations.[93] For models that conceptually overfit, benchmark-driven assessments are particularly convenient. Model validation focuses primarily on achieving equal or higher scores than previous models on the same datasets, tasks, and settings, effectively sidelining other important epistemic norms, such as conceptual accountability. Privacy nihilism wrongly equates high accuracy rates with model functionality, ignoring their lack of meaningful interpretation in conceptually overfitted models.

---

[85] Alternatively, the lower the better when assessing error rates.
[86] Kosinski M, Khambatta P and Wang Y, 'Facial Recognition Technology and Human Raters Can Predict Political Orientation from Images of Expressionless Faces Even When Controlling for Demographics and Self-Presentation' (2024) 79(7) *American Psychologist* 942-955.
[87] Gabriel Grill, 'Constructing Certainty in Machine Learning: On the Performativity of Testing and Its Hold on the Future' (2022).
[88] A standard set of metrics – for regressions, mean absolute error (MAE) or root mean squared error (RMSE), for example, or, for classification, accuracy, recall, specificity, and others – typically serves to demonstrate that an AI model performs a task as expected. This terminology owes its origin from AI development in the military, see: L McGuigan, *Selling the American People: Advertising, Optimization, and the Origins of Adtech* (MIT Press 2023).

In summary, conceptual overfitting refers to the adoption of norms of convenience that simplify the development of AI models by forcing complex constructs to fit data that are conceptually under-representative or irrelevant. Models built on an EfE approach conceptually overfit, conveniently inferring this from that at the expense of producing any meaningful conceptual accountability. Such processes typically fail to explicate why the collected data are relevant to the inference construct *in conceptual terms*. Consequently, practitioners circumvent a deliberate commitment to robust assumptions or claims, which would add friction to approaches that mandate looking for constructs of interests in data that one happens to have easy access to. A lack of conceptual accountability allows ground truth manufacturing to instantiate oversimplified constructs that can be easily attached to available data. Subsequently, model evaluation relies on accuracy scores as a heuristic, where "higher is better" becomes a rubber stamp for performance—disconnected from any meaningful sensemaking about the relationship between the inferred construct and the produced statistical results. Conceptual overfitting is not the absence of underlying concepts in EfE models; rather, it's the reluctance of developers and practitioners to critically evaluate foundational assumptions behind their constructs. Although convenient, this avoidance prevents conceptual accountability, making it impossible to understand and engage with the phenomena AI models truly infer.

---

[89] Birhane A, Kalluri P, Card D, Agnew W, Dotan R and Bao M, 'The Values Encoded in Machine Learning Research' (2022) *Proceedings of the 2022 ACM Conference on Fairness, Accountability, and Transparency* 173
[90] Raji ID, Bender EM, Paullada A, Denton E and Hanna A, 'AI and the Everything in the Whole Wide World Benchmark' (2021) *arXiv preprint* https://arxiv.org/abs/2111.15366 accessed 29 November 2024; Raji ID, Kumar IE, Horowitz A and Selbst A, 'The Fallacy of AI Functionality' (2022) *Proceedings of the 2022 ACM Conference on Fairness, Accountability, and Transparency* 959 https://doi.org/10.1145/3531146.3533150 accessed 29 November 2024; Church KW and Kordoni V, 'Emerging Trends: SOTA-Chasing' (2022) 28(2) *Natural Language Engineering* 249 https://doi.org/10.1017/S1351324921000374 accessed 29 November 2024; Grill, *Constructing Certainty in Machine Learning* (2022).
[91] Church and Kordoni, *Emerging Trends: SOTA-Chasing* (2022)
[92] Thomas RL and Uminsky D, 'Reliance on Metrics is a Fundamental Challenge for AI' (2022) 3(5) *Patterns*.
[93] Raji et al., *AI and the Everything in the Whole Wide World Benchmark* (2021)

**IV) AI CHALLENGES TO PRIVACY: THE PERSISTENT INFERENCE PROBLEM**

Taking stock of the argument to this point, conceptual overfitting is an antidote insofar as it challenges the underpinnings of privacy nihilism. Instead of substantiating the radical proposition that AI yields the power to infer everything from everything (EfE), many of the most dramatic, rhetorically persuasive displays of AI's boundless capacity, are epistemically flawed, potentially verging on epistemic arbitrariness. These flaws are due to, what we have called, conceptual overfitting, permeating key components of inference-making—data collection (or creation), ground truth manufacturing, and model evaluation. In revealing these flaws, our aim has been not only to undermine the foundations of privacy nihilism, but to stanch a retreat from privacy advocacy through regulation that a nihilistic position would suggest: Discarding data categories as a normative anchoring in privacy and data protection as a result of an *unconditional acceptance* of AI's inferential capacities. At the same time, as successful as we may be at achieving this aim, there is no denying that AI's power to amplify inferential capacity poses unprecedented challenges to privacy regulation—even if not to the point of privacy nihilism's EfE.

The most undermined by AI's power to amplify inferential capacity are approaches to privacy regulation that rely on distinctions among various types of data, traditionally, a dichotomy that marks certain categories of data as sensitive, private, or personal and according them special consideration. For example, regulation based privacy defined as a right to control information about ourselves might extend only to sensitive information, or, where it is defined as a form of secrecy, may extend only to these special classes of information.[94] On the other side of this dichotomy, either implied, or explicitly stated, is its opposite, namely, data that is "public"—neither sensitive nor private, privacy—to which regulation does not apply, or is muted. Such approaches are unquestionably challenged by powerful inferential capacities of AI, to the extent they enable private information to be inferred from its obverse.[95]

**INTRODUCING CONTEXTUAL INTEGRITY: A BRIEF DIGRESSION**

Developed in response to this and other challenges to privacy posed by digital technology, contextual integrity (CI) addresses gaps in other conceptions by positing contextual factors relevant to privacy that others had not systematically theorized.[96] At first glance, however, it doesn't seem to save CI from the inference attacks to which these alternative approaches are vulnerable. Although, on the one hand, CI demands finer-grained, data ontologies, rooted in social contexts (i.e. not public/private but, for example, symptoms, diagnoses, drugs prescribed, etc.) the data flows that preserve contextual integrity (i.e. are *appropriate*) still may yield inappropriate data flows due to clever data or model driven inferences. One may reasonably wonder whether there is a path forward for a meaningful conception of privacy (including those embodied in omnibus privacy regulation noted above) to resist radical inference-making.

The lens of contextual integrity can offer some clarity on the problem space and what is at stake in formulating a response.[97] To begin, CI rests on an assumption about social life drawn from disciplines of social theory and social philosophy, namely, that it is constituted by distinctive social domains, or realms, such as, healthcare, education, political governance, commerce, each defined by respective ends, purposes, and values. These contexts—for this is the term-of-art adopted in CI theory—are further constituted by respective ontologies of roles and data types. Behavior is governed by contextual norms, which may or may not be rendered as explicit rules or embedded in law and other forms of policy and regulation. A subset, contextual informational norms (abbreviated as, privacy norms), govern information flows by prescribing or proscribing them.[98] Appropriateness of information flows, at the heart of contextual integrity, is a function of privacy norms.

---

[94] This language is common in discussions of differential privacy.
[95] Helen Nissenbaum, 'Protecting Privacy in an Information Age: The Problem of Privacy in Public' in *The Ethics of Information Technologies* (Routledge 2020) 141-178: "*Information belies the adage about sewing silk purses out of sow's ears, for out of worthless bits of information we may sew assemblages that are rich in value.*"
[96] Helen Nissenbaum, 'Privacy as Contextual Integrity' (2004) 79 Washington Law Review 119.
[97] A few key aspects of CI theory are important to mention here, though interested readers may wish to seek fuller elaboration, explanation, and argument in other publications. For example: Helen Nissenbaum, *Privacy in Context: Technology, Policy, and the Integrity of Social Life* (Stanford University Press 2009).

The top level statement of CI, namely, that privacy is appropriate information flow draws into play all the elements we have introduced thus far. There are a few more, however, that we need to introduce. First, privacy norms comprise five parameters: subject, sender, recipient, data type (attribute), and transmission principle. When expressing a privacy norm, all five parameters must be specified and they must be specified in terms of the capacities (roles) in which the actors are acting and for the data types, in terms of respective contextual ontologies (for this reason, CI is a semantic theory). The transmission principle parameter refers to constraints under which data flows from one party to another, most familiarly, "with consent," but, potentially, innumerably others, such as, "reciprocally," "with authorization," "with notice," and so forth. When evaluating the privacy standing of an existing socio-technical practice, or one that is in development—be it ad targeting, facial recognition, or a myriad other possibilities—a description of any of the associated data flows must include values for all of the parameters. Failing to do so renders it incomplete and ambiguous. Second, privacy norms inherit an ambiguity that is not unusual when characterizing social norms, namely, they may serve as descriptions of common or conventional behaviors and expectations, or may be expressed as morally weighty prescriptions, or both, simultaneously. In either case, CI introduces the idea of legitimacy: privacy norms are legitimate, or, morally defensible, insofar as they support contextual ends, purposes, and values and illegitimate, or morally indefensible, insofar as they undermine them.[99]

---

[98] Because it isn't the case that there are norms governing all possible information flows, it is possible that for a given data flow, the question of its appropriateness is indeterminate.
[99] The same may be said of data flows, even in the absence of entrenched norms, viz. that they are legitimate or illegitimate.

The inference challenge, through the lens of CI, is most acute for conceptions of privacy that rest entirely, or almost entirely, on the single parameter of attribute, or data type without qualifying, further, the values of other—ideally, all the other—parameters required for flows to be appropriate. Although this may seem overly pedantic, it is worth noting that failure to do so has allowed data brokers, for decades, to justify their unremitting practices of seeding massive individual dossiers by scraping information from so-called public records. This failure has also served companies, such as, OpenAI and ClearviewAI, who have defended their data practice by declaring that they collect only "publicly available" data.[100] Even without incorporating missing parameters, one immediate approach, might be to adjust the categories of private and public by switching data attributes to "private" that previously had been classified as "public," as occurred with the Social Security Number in the US. At most, however, this can serve as a stopgap, inevitably outpaced by potent advancements in AI based inference-making.[101]

CI opens up an array of effective alternatives for regulation, allowing that all five parameters are in play as levers of governance. Not as radical as this might initially seem, there is already law on the books that has done this, HIPAA, for example, which spells out the proper recipients of different types of health information, acting in their respective capacities, including physicians, mental health professionals, business services, and so forth. The HIPAA privacy rules spell out not only the proper recipients but also what the recipients may and may not do with information received and under what conditions. Academic researchers pledge to human subjects that specified restrictions will curtail how the researchers handle data collected from and about them. Regulation curbs not only *how* law enforcement agents (in liberal democracies) may go about gathering information about suspects but also what that information may include. The last, illustrates the power of transmission principles to stipulate prior conditions and constraints on data flows, even when the other parameters are non-determinative for many actions. In some cases, when parties in an instance of data flow are acting in historically familiar capacities, for example, teachers or lawyers, their obligations, determined exogenously, may be imported into entrenched norms governing it. In sum, giving regulators

---

[100] These declarations were made in respective Privacy Policies, slightly more complicated in the case of OpenAI, who admit that they also obtain data through licensing agreements with other data holders.
[101] See discussion in H. Nissenbaum (2019) ["Contextual Integrity Up and Down the Data Food Chain,"](#) *Theoretical Inquiries in Law 20:1*, 221-256

(governmental and nongovernmental) the discretion to govern data practices through all of the five parameters, not merely data type, means, at least, that when data is generated in individuals' interactions with others, including, business, services, platforms, apps, etc. they will be held accountable for the capacities in which they are acting and like lawyers and teachers, held to specific restrictions.

A fuller account of how the CI parameters could provide critical levers to privacy regulation would require more painstaking detail than this article allows. Before concluding this section, however, it is critical to mention that tailoring data flow norms to sensibly limit the application of AI's inferential power is no arbitrary exercise. It is not about simply capping, reducing, and stopping; instead, legitimacy remains the guiding light. Just as carefully crafted rules governing data flows in healthcare are justifiable only if they promote its purposes and values, norms governing data flows in the provision of legal services promote provision by lawyers of individuals rights to representation, and desired norms in education would promote the ends of learning so must we evaluate restrictions on nontraditional data practices. If societal and contextual values are disserved by existing data flows, any new normative restrictions need to be evaluated against their capacity to turn the tables.

## V) CONCLUSION

By centering privacy exclusively on data types—only to abandon them by granting AI models unconditional inferential power—privacy nihilism fails to offer any substantive advancement for privacy in a world shaped by ubiquitous AI inference-making. In this article, we have shown that while AI may *seem* to transcend any data category and possibly infer "everything from everything," such claims rest on fragile epistemic foundations. We introduced the notion of conceptual overfitting to describe how AI model development can follow norms of convenience, mapping complex constructs—including sensitive data types—onto readily available data that is often at hand but conceptually under-representative or even irrelevant. It may be more surprising, in fact, that the dominant data economy allows for "anything goes" in acquiring, generating, and capturing data and for utilizing it at the arbitrary convenience of the burgeoning array of actors—commercial but not always—who either accrue data and generate data products as their core business or who generate (collect, capture) data as an incidental artifact of their primary service. We documented how conceptual overfitting can arise when requesters instruct labeling of data as if it were adequately capturing rich, nuanced constructs, such as political orientation, mental health status, or religious beliefs. Accuracy metrics can offer a veneer of objectivity that masks the absence of genuine conceptual grounding—leading to a false impression that AI models have attained near-limitless inference capabilities. In summary, we have argued that "everything from everything" perspectives fail to provide a justified descriptive foundation from which to derive normative conclusions about privacy in both theory and law. Crucially, our findings do not refute the mounting challenges that novel AI systems pose for privacy regulation. Instead of calling for the outright dismissal of data categories, we have underscored the importance of other factors, such as those proffered in contextual integrity, including, social norms, key actors, and contextual purposes and values. By factoring in these additional elements, privacy values are not relegated to the back but remain central to privacy protection as AI's inferential capabilities increase. An epistemic "anything goes" approach to AI inferences disrupts the equilibrium that an adequate theory of privacy would seek to reach. Why should the authority to determine the legitimacy of inferences rest solely with those who produce, develop, and deploy inferential AI models? Who stands to gain, and who stands to lose, when such positions are advanced? These questions lie at the heart of the possibilities and limitations of governing information in the age of AI inferences. Instead of giving in either to privacy dogmatism or to privacy nihilism, defending privacy will mean avoiding hyperbole while taking a clear-eyed approach to the real challenges of AI's dramatic powers of inference-making.